\documentclass[10pt]{article}
\usepackage{epsfig}

\makeatletter
\renewcommand\section{\@startsection {section}{1}{\z@}%
                                   {-3.5ex \@plus -1ex \@minus -.2ex}
                                   {2.3ex \@plus.2ex}%
                                   {\normalfont\large\bfseries}}
\renewcommand\subsection{\@startsection{subsection}{2}{\z@}%
                                     {-3.25ex\@plus -1ex \@minus -.2ex}%
                                     {1.5ex \@plus .2ex}%
                                     {\normalfont\bfseries}}
\makeatother

\long\def\symbolfootnote[#1]#2{\begingroup%
\def\thefootnote{\fnsymbol{footnote}}\footnote[#1]{#2}\endgroup}
\newcommand{\be}{\begin{equation}}
\newcommand{\ee}{\end{equation}}
\newcommand{\ber}{\begin{array}}
\newcommand{\eer}{\end{array}}

\newcommand{\del}{\partial}

\newcommand{\dsty}{\displaystyle}

\newcommand{\de}{\delta}

\newcommand{\eps}{\varepsilon}

\newcommand{\br}[1]{\left(#1\right)}

\begin{document}

\begin{flushright}
arXiv:yymm.nnnn [hep-th]
\end{flushright}\vspace{5mm}

\begin{center}
{\LARGE\bf Notes on quantum evolution across singularities}    \\
\vskip 5mm
{\large Oleg Evnin}
\vskip 3mm
{\em Theoretische Natuurkunde, Vrije Universiteit Brussel and\\
The International Solvay Institutes\\ Pleinlaan 2, B-1050 Brussels, Belgium}
\vskip 2mm
{\small\noindent  {\tt eoe@tena4.vub.ac.be}}\end{center}\vspace{8mm}

\noindent {\small In a number of model contexts, evolution across space-time singularities
(reminiscent of the cosmological singularities) involves time-dependent
quantum Hamiltonians developing a singularity as a function of time.
In this contribution to the proceedings of the 3rd RTN workshop in Valencia, I review
some recent investigations of the general properties of such systems.}\vspace{8mm}

\begin{flushright}
\small\it ``Ecce respondeo dicenti: `Quid faciebat Deus, antequam faceret caelum et terram?'\\ Respondeo non illud, quod quidam respondisse perhibetur ioculariter eludens\\ quaestionis violentiam: `Alta, inquit, scrutantibus gehennas parabat.''\vspace{1mm}\\
Augustini Confessiones, Liber XI, Caput XII\symbolfootnote[2]{``How shall I respond to him who asks, `What was God doing before he made heaven and earth?' I shall not respond as that other person who, avoiding the power of the question, replied in a joke: `He was preparing hell for those who research too deep.'' St. Augustine's Confessions, Book XI, Chapter XII}
\end{flushright}

\section{Introduction}

The interest for dynamical transitions through space-time singularities has been revived in the recent years: both in relation to the ekpyrotic cosmological scenario (see \cite{ekp} and the subsequent work) and as a formal problem in quantum space-time studies:
in the context of pertutbative string theory (\cite{lms} and the subsequent work), as
well as its extensions (\cite{benreview} and references therein; see also \cite{ads}).

From the standpoint of potential cosmological applications, it is, of course, space-like singularities that one should be interested in most. Such singularities undermine the validity of the standard initial value problem and demand further specifications for treating the singular region. Of particular interest is the question whether the dynamical evolution can be extended across the singularity.

Given our presently limited theoretical understanding of quantum gravity, studying space-like singularities often appears to be unviable. Light-like singularities
present an interesting alternative in that they introduce an obstruction to
the conventional dynamical evolution in a way very similar to space-like singularities,
but often appear more formally tractable. Furthermore, the Penrose limit converts
space-like singularities into light-like singularities, though the precise implications
of this formal observation are unknown.

In a number of model contexts, quantum evolution across space-time
singularities appears to be described by time-dependent Hamiltonians
developing an isolated singularity as a function of time at the moment the system
reaches a space-time singularity. Needless to say, additional
specifications are needed in a Schr\"odinger equation involving this
kind of Hamiltonians, on account of the singular time dependence.

One of the simplest examples of such singular time-dependent Hamiltonians in systems
with space-time singularities is given by a free scalar field
on the Milne orbifold. Because the squareroot determinant of the metric of the Milne orbifold $ds^2=-dt^2+t^2dx^2$
vanishes as $|t|$ when $t$ goes to 0,
the kinetic term in the Lagrangian for a free field $\phi$
on the Milne orbifold will have the form $|t|(\del_t\phi)^2$
and the corresponding term in the Hamiltonian expressed
through the canonical momentum $\pi_\phi$ conjugate to $\phi$
will have the form $\pi_\phi^2/|t|$,
which manifestly displays an $1/|t|$ singularity. The position of this
singularity in the time dependence
coincides with the metric singularity of the Milne orbifold.

While it is well-known that free fields on the Milne orbifold are not a good approximation
to interacting systems, especially in gravitational theories,
analogous singular time dependences have recently appeared in other models,
which have been the main motivation for the present work. For example,
11-dimensional quantum gravity with one compact dimension
in a certain singular time-dependent background with a light-like
isometry is conjectured to be described by a time-dependent
modification of matrix string theory \cite{ben}.
This model can be recast in the form of a (1+1)-dimensional
super-Yang-Mills theory on the Milne orbifold. It will thus contain
in its Hamiltonian the $1/|t|$ time dependence typical of the general Milne orbifold
kinematics. The question of transition through the singularity will
then amount to defining a quantum system with such singular Hamiltonian.
Likewise, for the time-dependent matrix models of \cite{miaoli}, which
are conjectured to describe quantum gravity in non-compact eleven-dimensional
time-dependent background with a light-like singularity, one obtains
a quantum-mechanical Hamiltonian with a singular time dependence.

In view of these observations, it appears worthwhile to study the general features
of quantum Hamiltonians with isolated singularities in their time dependence (we shall be considering resolved singularities, with a regularization parameter denoted as $\eps$, and subsequently take an $\eps\to 0$ limit). Most generally, one could simply say that the Hamiltonian is a time-dependent operator $H(t,\eps)$ developing a singularity, say, at $t=0$ when $\eps$ is taken to 0. This situation is hard to analyze, and we shall restrict ourselves to the case when the Hamiltonian is a combination of a few operator terms
with singular coefficients:
\be
H(t)=\sum\limits_i f_i(t,\eps)  H_i,
\label{multiop0}
\ee
where $H_i$ are time-independent operators and $f_i$ are time-dependent number-valued functions. $\eps$ is a regularization parameter, and the implication is that, as $\eps$ is taken to 0, some of the $f_i$'s may develop isolated singularities at a certain value of $t$, which we shall choose to be $t=0$. Our ultimate question will be whether the $\eps\to 0$ limit of the evolution operator corresponding to (\ref{multiop0}) exists (this is not generally so, since singularities in $H(t)$ would typically introduce non-integrable singularities in the Schr\"odinger equation).

Given a quantum Hamiltonian with a singularity in its time dependence at $t=0$, there are
two issues one needs to address. Firstly, one needs to construct a regularization (\ref{multiop0}) in such a way that: (a) the $\eps\to 0$ limit of the evolution operator exists, (b) the $\eps\to 0$ limit of (\ref{multiop0}) equals the original singular Hamiltonian away from $t=0$. Secondly, one should decide which such regularization to choose.

The latter question is very subtle, since, if all one demands is that the evolution
away from $t=0$ is the same as the one given by the original (singular) Hamiltonian and the evolution operator is unitary,
one can insert an arbitrary unitary transformation at $t=0$, and the predictive power is lost completely. One therefore needs to formulate some qualitative restrictions
on the regularization procedure, which would choose a particular form of (\ref{multiop0})
for the given (singular) Hamiltonian.

In the absence of further physical specifications, it is natural to adopt a ``minimal subtraction'' approach \cite{qsing}. Namely, one can try to regulate a singular Hamiltonian without introducing any additional operator structures. This is the simplest possible prescription, and I shall start by reviewing it below.

In the context of geometrical theories, it is natural to demand that the regularizations
(\ref{multiop0}) admit a geometrical interpretation at any value of $\eps$ (or, even stronger, satisfy Einstein's equations or other such background consistency conditions).
This approach typically takes one outside of the context of the ``minimal subtraction'' recipe and enforces regularizations (\ref{multiop0}), where multiple operator structures essentially contribute in the singular region. This will be the second topic reviewed in these notes.

\section{Minimal subtraction and the Milne orbifold}

Following the general remarks in the introduction, we shall consider
a quantum system described by the following time-dependent Hamiltonian:
\be
H(t)=f(t,\eps){h}+H_{reg}(t),
\ee
where $H_{reg}(t)$ is non-singular around $t=0$, whereas the numerical function $f(t,\eps)$ develops an isolated singularity at $t=0$ when $\eps$ goes to 0 ($\eps$ serves as a singularity regularization parameter), and $h$ is a time-independent operator. We shall be interested in the
evolution operator from small negative to small positive time. In this region,
we shall assume that we can neglect the regular part of the Hamiltonian $H_{reg}(t)$
compared to the singular part.%
\footnote{Generally, this doesn't need to be the case. Whenever it isn't, a full treatment along the lines of section 3 should be applied.}
The Schr\"odinger equation takes the form
\be
i\frac{d}{dt}|\Psi\rangle=f(t,\eps)h|\Psi\rangle.
\label{schr}
\ee
And the solution for the corresponding evolution operator is obviously given by
\be
U(t,t')=\exp\left[-i\int\limits_t^{t'} dt f(t,\eps) h\right].
\label{ev}
\ee
When the regularization parameter $\eps$ is sent to 0, $f(t,\eps)$ becomes singular 
and $U(t,t')$ is in general not well-defined.

The goal is then to modify the Hamiltonian locally at $t=0$ in such a way that the evolution away from $t=0$ remains as it was before, but there is a unitary transition
through $t=0$. Of course, a large amount of ambiguity is associated
with such a program, and we shall comment on it below.

The most conservative approach to the Hamiltonian modification is suggested by
(\ref{ev}). Since the problem arises due to the impossibility of integrating
$f(t,\eps)$ over $t$ at $\eps=0$, the natural solution is to modify $f(t,\eps)$
locally around (in the $\eps$-neighborhood of) $t=0$ in such a way that the
integral can be taken (note that we are leaving the operator structure of the 
Hamiltonian intact).

The subtractions necessary to appropriately modify $f(t,\eps)$ are familiar
from the theory of distributions. Namely, for any function $f(t,\eps)$
developing a singularity not stronger than $1/t^p$ as $\eps$
is sent to 0, with an appropriate choice of $c_n(\eps)$, one can introduce a modified
\be
\tilde f(t,\eps)=f(t,\eps)-\sum\limits_{n=0}^{p-1}c_n(\eps)\de^{(p)}(t)
\label{distr_subtr}
\ee
(where $\de^{(p)}(t)$ are derivatives of the $\de$-function) in such a way
that the $\eps\to 0$ limit of $\tilde f(t,\eps)$ is defined in the sense
of distributions. The latter assertion would imply that the $\eps\to 0$ limit of 
\be\int \tilde f(t,\eps) {\cal F}(t) dt\ee 
is defined for any smooth ``test-function'' ${\cal F}(t)$, and, in particular, that the $\eps\to 0$ limit of (\ref{ev}) becomes well-defined,
if $f(t,\eps)$ is replaced by $\tilde f(t,\eps)$. (Note that, since $f(t,\eps)$ and $\tilde f(t,\eps)$ only differ in an infenitesimal neighborhood of $t=0$, this modification will
not affect the evolution at finite $t$).

As a matter of fact, the subtraction needed for our particular case is simpler that
(\ref{distr_subtr}). Since the $n>0$ terms in (\ref{distr_subtr}) can only affect
the value of the evolution operator (\ref{ev}) at $t'=0$, if one is only interested
in the values of the wave function for non-zero times, one can simply omit the $n>0$ terms from (\ref{distr_subtr}). One can then write down the subtraction explicitly as
\be
\tilde f(t,\eps)=f(t,\eps)-\left(\int\limits_{-t_0}^{t_0} f(t,\eps) dt\right)\de(t).
\ee
The appearance of a free numerical parameter (which can be chosen as $t_0$ in the expression above, or a function thereof) is not surprising, since, if $\tilde f(t,\eps)$ is an adequate modification of $f(t,\eps)$, so is $\tilde f(t,\eps) + c\de(t)$ with any finite $c$.

One should note that it is very natural to think of the above subtraction procedure
as renormalizing the singular time dependence of the Hamiltonian. Indeed the mathematical
structure behind generating distributions by means of $\de$-function subtructions is
precisely the same as the one associated with subtracting local counter-terms
in order to render conventional field theories finite. To make this analogy more
transparent, one can briefly envisage the computation of the following
simple diagram in the familiar $\lambda\phi^3$ field theory:
\begin{center}
\begin{picture}(150,42)
\put(0,0){\epsfig{file=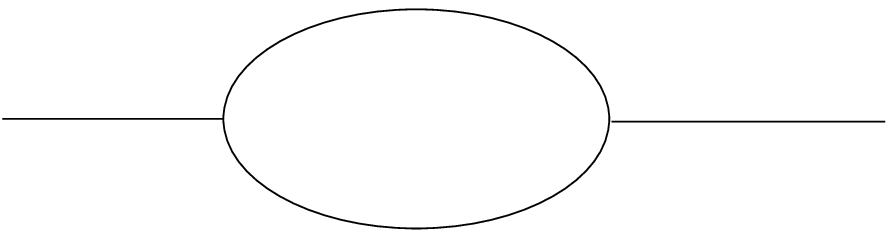,width=5cm}}
\put(28,8){$x$}
\put(105,7){$x'$}
\end{picture}
\end{center}
If evaluated in position (rather than in momentum) space, the expression for this
diagram will not contain any integrations, so one might wonder where the familiar
divergences would come from. The problem is that the diagram contains the
square of the scalar field propagator $[D(x,x')]^2$, and, whereas the propagator
itself if a distribution, its square is not. For that reason, if one tries,
for example, to evaluate a Fourier transform (momentum space expression) for this
diagram, one obtains infinities, since integrals of $[D(x,x')]^2$ cannot be evaluated.
The problem is resolved by subtacting local counter-terms from the field theory
Lagrangian, which, for the above diagram, would translate into adding $\de(x-x')$
and its derivatives (with divergent cutoff-dependent coefficients) to $[D(x,x')]^2$ in such a way
as to make it a distribution. The mathematical structure of this procedure is
precisely the same as what we employed for renormalizing the singular time
dependences in time-dependent Hamiltonians.

As an application of these ideas, one can consider (as a toy model) a free scalar field on the Milne orbifold \cite{qsing}. The metric is
\be\label{MilneMetric}
ds^2=-dt^2+t^2dx^2,
\ee
and the Hamiltonian is
\be
H=\frac1{2|t|}\int dx\,\left(\pi_\phi^2+{\phi^\prime}^2\right)+\frac{m^2|t|}2\int dx\, \phi^2.
\label{hmsingular}
\ee
For the particular $1/|t|$ time dependence featured in (\ref{hmsingular}), one can choose
$f(t,\eps)$ as $1/\sqrt{t^2+\eps^2}$, in which case $\tilde f(t,\eps)$ becomes
\be
f_{1/|t|}(t,\eps)=\frac{1}{\sqrt{t^2+\eps^2}}+2\ln(\mu\eps)\de(t),
\ee
or more accurately (with an $\eps$-resolved $\de$-function)
\be
f_{1/|t|}(t,\eps)=\frac{1}{\sqrt{t^2+\eps^2}}+2\ln(\mu\eps)\frac{\eps}{\pi(t^2+\eps^2)}.
\label{f1/t}
\ee
(with $\mu$ being an arbitrary mass scale). The regulated Hamiltonian then reads:
\be
H=\frac1{2} f_{1/|t|}(t,\eps)\int dx\,\left(\pi_\phi^2+{\phi^\prime}^2\right)+\cdots.
\label{hmreg}
\ee
As explained in \cite{qsing}, this ``minimal subtraction'' treatment is essentially the same as the covering Minkowski space recipes previously proposed in the literature for this situation (even though those recipes correspond to different values of the parameter $\mu$ in (\ref{f1/t}) for the different oscillator modes of the scalar field).

We should remark upon the general status of our Hamiltonian prescription
viewed against the background of all possible singularity transition recipes one could devise.
If the only restriction is that the evolution away from the singularity is
given by the original Hamiltonian, one is left with a tremendous infinitefold ambiguity:
any unitary transformation can be inserted at $t=0$ and the predictive power is lost
completely. One should look for additional principles in order to be able to define
a meaningful notion of singularity transition.

Our prescription can be viewed as a very conservative approach, since it preserves
the operator structure of the Hamiltonian (the counter-terms added are themselves
proportional to $h$, the singular part of the Hamiltonian). In the absense of
further physical specification, this approach appears to be natural and can be
viewed as a sort of ``minimal subtraction''. However, under some circumstances,
one may be willing to pursue a broader range of opportunities for defining the singularity transition. For example, one may demand that the resolution of the singular dynamics must have a geometrical interpretation (at finite values of $\eps$).

\section{Geometrical resolutions, multiple operator structures\\ and the generalized null-brane}

The ``minimal subtraction'' approach of the previous section is very natural algebraically,
but, in a number of contexts, one may have additional qualitative principles constraining
the definition of the singularity transition. For example, if the original physics problem
features a (singular) space-time background, it may be natural to demand that its regularized version should also admit a geometrical interpretation (for all values of $\eps$). In string theory (or matrix model) contexts, it may be necessary to
demand that the regularized (geometrical) backgrounds satisfy some further consistency
conditions, such as Einstein's equations (or an appropriate generalization thereof).

Such additional requirements may well be in conflict with the ``minimal subtraction'' ansatz.
For example, a direct inspection of (\ref{hmreg}) shows that the regularized version of the scalar field dynamics does not admit a geometrical interpretation (nor should one think of its singular limit, albeit well-defined, as being geometrical).

The problem with constructing a geometrical interpretation of (\ref{hmreg}) is that, since $f_{1/|t|}(t,\eps)$ has an $\eps\to 0$ limit as a distribution, the $\eps\to 0$ limit of
\be
\int\limits_{-t_0}^{t_0}dt\,f_{1/|t|}(t,\eps)
\label{int}
\ee
must exist. Furthermore, as stated above, the $\eps\to 0$ limit of $f_{1/|t|}(t,\eps)$ must equal $1/|t|$ everywhere away from $t=0$. For that reason, in order for the limit of (\ref{int}) to exist, $f_{1/|t|}(t,\eps)$ should be very large and {\it negative} somewhere in the $\eps$-neighborhood of $t=0$ so that the positive divergence from integrating $1/|t|$ is compensated in (\ref{int}). However, the coefficient of the kinetic term in the Hamiltonian of a field in a geometrical background comes from the square root of the determinant of the metric (and the coefficients of the inverse metric), and it needs to be {\it positive} (as is the function $1/|t|$ appearing in (\ref{hmsingular})).

For that reason, there appears to be a conflict between the demands of positivity for certain coefficients in the Hamiltonian arising if one pursues a geometrical interpretation, and negative contributions introduced by our ``minimal subtraction'' recipe. If one is to construct a geometrical resolution of dynamics on a singular space-time background, one generally needs to relax the specifications of the ``minimal subtraction'' approach, and permit modifications in the operator structure of the Hamiltonian, as well as its time dependence, in the vicinity of the singular region. One will then typically end up with a situation where a few different operator structures in the Hamiltonian essentially contribute to the transition in the singular region:
\be
H(t)=\sum\limits_i f_i(t,\eps)  H_i,
\label{multiop}
\ee
It is the commutation properties of the different terms in the Hamiltonian that are responsible for divergence cancellation (rather than explicit negative contributions introduced through the ``minimal subtraction'' scheme of \cite{qsing}).

It is in general impossible to solve the Schr\"odinger equation corresponding to the Hamiltonian (\ref{multiop}). The familiar symbolic solution for the evolution operator $U(t_1,t_2)$ involves the time-ordering symbol ${\mbox T}$:
\be
U(t_1,t_2)={\mbox T}\left[-i\int\limits_{t_1}^{t_2} dt\, H(t)\right].
\label{timeorder}
\ee
Our principal question is whether the $\eps\to 0$ limit of this evolution operator exists.

The above representation can be further transformed in an instructive way using a technique known as the Magnus expansion \cite{magnus}. The operator $U$ belongs to the group of unitary operators on the Hilbert space, and the Magnus expansion can be thought of as an analog of the Baker-Campbell-Hausdorff formula for finite-dimensional Lie groups. The expansion can be symbolically written as:
\be
\begin{array}{l}
\dsty U(t_1,t_2)=\exp\left[-i\int\limits_{t_1}^{t_2}dt\,H(t)+\eta_1\int dt\,dt'\,[H(t),H(t')]\right.\vspace{2mm}\\
\dsty\hspace{4cm}\left.+i\,\eta_2
\int dt\,dt'\,dt''\,[H(t),[H(t'),H(t'')]]+\cdots\right],
\end{array}
\label{magnus}
\ee
with some numerical coefficients $\eta_1$, $\eta_2,\ldots$ (their values will not be important for us, and it appears they can only be derived recursively \cite{magnus}). The key property of the above expression is that the higher order terms are entirely expressed through higher order nested commutators of $H(t)$ at different moments of time.

Even though, in a completely general setting, the Magnus expansion is hopelessly intractable, it displays the broad range of opportunities for divergence cancellation in a singular limit of the dynamics described by (\ref{multiop}). Namely, for the case of (\ref{multiop}), the Magnus expansion (\ref{magnus}) will contain all kinds of combinations of the $f_i$ and their products, in such a way that, even if $f_i$ develop very strong singularities as $\eps$ is taken to 0, the limit of $U(t_1,t_2)$ may still exist. For example, even if all $f_i$ are positive, cancellations may still occur on account of the commutation properties of $H_i$.

Should such cancellations take place, one may think of the $\eps\to 0$ limit of (\ref{multiop}) as an operator-valued generalization of conventional distributions: just as ordinary distributions may contain singularities in a way that permits evaluating ordinary integrals, the Hamiltonian (\ref{multiop}) will contain singularities in a way that permits evaluating the time-ordered exponential integral in (\ref{timeorder}). A systematic exploration of such generalized operator-valued ``distributions'' may be interesting to pursue, but lies outside of the scope of the present paper.

There is a special case when the above analysis can be taken significantly further. Namely, it may turn out that, for all moments of time, the operator $U$ of (\ref{timeorder})  belongs to a finite-dimensional subgroup of the unitary group of the Hilbert space. This situation has been described as a presence of a {\it dynamical group} (see \cite{zhengfenggilmore,malkinmanko} and references therein). For the Hamiltonians of the form (\ref{multiop}), there will exist a finite-dimensional dynamical group if the set of nested commutators of $H_i$'s closes on a finite-dimensional linear space of operators (which would serve as the Lie algebra of the dynamical group). Should that happen, one would be able to use the the closed resummed version of the Baker-Campbell-Hausdorff formula for finite-dimensional Lie groups to treat the Magnus expansion, or, alternatively, the Schr\"odinger equation can be reduced to a finite number of ordinary differential equations describing the evolution on the finite-dimensional dynamical group manifold \cite{zhengfenggilmore,malkinmanko}. In practical terms, one can choose a particular low-dimensional faithful linear representation of the dynamical group furnished by matrices $M$, and write down the Schr\"odinger equation in this representation:
\be
i\frac{dM(t,t_0)}{dt}=\varphi(H(t))M,\qquad M(t_0,t_0)=1,
\ee
where $\varphi$ is a homomorphism from Hilbert space operators onto the representation furnished by $M$.
(This is a finite-dimensional system of ordinary differential equations.) Given the solution for $M(t,t_0)$, one can reconstruct the original evolution operator as $\varphi^{-1}(M(t,t_0))$.

The case of a free scalar field on the generalized null-brane and its singular limit, the parabolic orbifold,
(see \cite{gres} for a complete consideration) falls precisely into this category.
I shall not reproduce here the relevant derivations, and simply state the results.
Further details can be found in \cite{gres}.

The generalized null-brane space time is described by the metric
\begin{equation}
ds^2=\frac{R^2 X^2 \br{\beta^2-\alpha}}{\br{R^2+(X^+)^2}^2}\br{dX^+}^2\,-2 dX^+ dX^-+ \frac{2\beta R X}{\sqrt{R^2+(X^+)^2}}dX^+ d\Theta+\br{R^2+(X^+)^2} d\Theta^2+dX^2,\label{lineElementGNB}
\end{equation}
where $\alpha$, $\beta$ and $R$ are parameters. The case $\alpha=3, \beta=2$ corresponds to the original null-brane. And the $R\to 0$ limit (irrespectively of the values of $\alpha$ and $\beta$) is the (singular) parabolic orbifold:
\begin{equation}
ds^2=-2dX^+dX^-+(X^+)^2 d\Theta^2\label{metricRosenPO}
\end{equation}
(times a line parametrized by $X$).

For a free scalar field on the generalized null-brane space time, one finds a
finite-dimensional dynamical group structure (with the so-called ``two-photon group''),
which essentially reduces the dynamics to a single ordinary differential equation.
Eventually, one finds that the $R\to 0$ limit of the scalar field dynamics exists
if and only if $\alpha$ and $\beta$ belong to the following discrete spectrum:
\be
\alpha=(2N)^2-1,\qquad \beta=2N,
\ee
with $N$ being an integer.

The appearance of discreteness is quite surprising in this context, and it may have interesting qualitative consequences if it persists to more realistic models featuring space-time singularities.

\section{Conclusions}

I have reviewed some recent investigations of singularities in the time dependence of quantum Hamiltonians. This type of systems has appeared in a few different contexts in relation to studies of space-time singularities.

It has been emphasised in this setting that defining a dynamical transition through
the singular point involves a large amount of ambiguity. One approach is that of algebraic simplicity, which leads to the ``minimal subtraction'' recipe described in the text.

Demanding singularity resolutions with a geometrical interpretation tends to take
one outside the constraints of the ``minimal subtraction'' method. One is then forced
to consider Hamiltonians whose near-singular time dependence essentialy contains multiple operator structures. Analyzing the singular limit of such Hamiltonians involves some interesting and non-trivial mathematical problems. In a particular example considered here
(that of a free scalar field on the generalized null-brane space-time), one discoveres rather intriguing discrete features of the singular limit.

\section{Acknowledgments}

I would like to thank Ben Craps and Frederik De Roo for collaboration on the subjects presented in this note. This research has been supported in part by the Belgian Federal Science Policy Office through the Interuniversity Attraction Pole IAP VI/11, by the European Commission FP6 RTN programme MRTN-CT-2004-005104 and by FWO-Vlaanderen through project G.0428.06.

\end{document}